\documentclass[onecolumn,amsmath,showpacs,nofootinbib,11pt]{revtex4-1}
\usepackage{graphicx}
\usepackage{dcolumn}
\usepackage{bm}
\begin{document}
\newcommand{\hs}{\hspace*{0.5cm}}
\newcommand{\vs}{\vspace*{0.5cm}}
\newcommand{\be}{\begin{equation}}
\newcommand{\ee}{\end{equation}}
\newcommand{\bea}{\begin{eqnarray}}
\newcommand{\eea}{\end{eqnarray}}
\newcommand{\ben}{\begin{enumerate}}
\newcommand{\een}{\end{enumerate}}
\newcommand{\bde}{\begin{widetext}}
\newcommand{\ede}{\end{widetext}}
\newcommand{\nn}{\nonumber}
\newcommand{\crn}{\nonumber \\}
\newcommand{\Tr}{\mathrm{Tr}}
\newcommand{\non}{\nonumber}
\newcommand{\noi}{\noindent}
\newcommand{\al}{\alpha}
\newcommand{\la}{\lambda}
\newcommand{\bet}{\beta}
\newcommand{\ga}{\gamma}
\newcommand{\va}{\varphi}
\newcommand{\om}{\omega}
\newcommand{\pa}{\partial}
\newcommand{\+}{\dagger}
\newcommand{\fr}{\frac}
\newcommand{\bc}{\begin{center}}
\newcommand{\ec}{\end{center}}
\newcommand{\Ga}{\Gamma}
\newcommand{\de}{\delta}
\newcommand{\De}{\Delta}
\newcommand{\ep}{\epsilon}
\newcommand{\varep}{\varepsilon}
\newcommand{\ka}{\kappa}
\newcommand{\La}{\Lambda}
\newcommand{\si}{\sigma}
\newcommand{\Si}{\Sigma}
\newcommand{\ta}{\tau}
\newcommand{\up}{\upsilon}
\newcommand{\Up}{\Upsilon}
\newcommand{\ze}{\zeta}
\newcommand{\ps}{\psi}
\newcommand{\Ps}{\Psi}
\newcommand{\ph}{\phi}
\newcommand{\vph}{\varphi}
\newcommand{\Ph}{\Phi}
\newcommand{\Om}{\Omega}
\newcommand{\AdrHEPC}{Phenikaa Institute for Advanced Study and Faculty of Basic Science, Phenikaa University, Yen Nghia, Ha Dong, Hanoi 100000, Vietnam}

\title{Gauge origin of double dark parity and implication for dark matter}

\author{Duong Van Loi}
\email{loi.duongvan@phenikaa-uni.edu.vn} 
\author{Phung Van Dong} 
\email{dong.phungvan@phenikaa-uni.edu.vn}
\affiliation{\AdrHEPC} 

\date{\today}

\begin{abstract}
Dark matter must be stabilized over the cosmological timescale, which demands the existence of a stabilizing symmetry, derived by a dark charge, $D$. The existence of this dark charge may affect the quantization of electric charge, which theoretically shifts the electric charge, thus the hypercharge to a novel gauge extension, $SU(3)_C\otimes SU(2)_L\otimes U(1)_Y\otimes U(1)_N$, where $N$ determines $D=T_3+N$, similar to $Q=T_3+Y$. New observation of this work is that the dark charge is broken down to two kinds of dark parity, $Z_2$ and $Z'_2$, which subsequently imply three scenarios of dark matter. The relic density and direct detection for the scenario of two-component dark matter are investigated in detail.     
\end{abstract}


\maketitle

\section{Introduction}

The detection experiments of neutrino oscillations have shown that the neutrinos have nonzero small masses and flavor mixing, which cannot be addressed within the framework of the standard model \cite{RevModPhys.88.030501,RevModPhys.88.030502}. Furthermore, the standard model fails to explain dark matter, which makes up most of the mass of the galaxies and galaxy clusters \cite{Hinshaw:2012aka,Aghanim:2018eyx}.     

Vast attempts have been paid to solve these long-standing questions, basically given in terms of a seesaw or/and radiative mechanism \cite{Minkowski:1977sc,GellMann:1980vs,Yanagida:1979as,Glashow:1979nm,Schechter:1980gr,Weinberg:1979sa,Zee:1980ai,Zee:1985id,Babu:1988ki} to induce a small neutrino mass, with implement of a discrete symmetry \cite{Krauss:2002px,Ma:2006km,Okada:2010wd} to stabilize a dark matter candidate. The small neutrino masses can be appropriately generated by such a mechanism, but the existence and stability of dark matter is {\it ad hoc} introduced. The discrete symmetry that stabilizes the dark matter is eventually a $Z_2$ or a matter parity as in supersymmetry \cite{Martin:1997ns}, which is not naturally conserved by the theory.     

It is shown that the dark matter stability symmetry may relax the quantization of electric charge, similar to anomaly-free hidden symmetries studied in \cite{Babu:1989tq,Babu:1989ex,Foot:1990uf}. Therefore, a dequantization version of electric charge that deviates from the electric charge by a parameter already defines a dark charge, i.e. a dark symmetry, to be a novel gauge extension \cite{VanDong:2020cjf}. The dark charge breaking not only produces the neutrino masses and makes the electric charge quantized, but also determines a dark parity as residual gauge symmetry, which is different from the studied $Z_2$/matter parity and not commuted with the weak isospin, similar to the electric charge. This approach provides a possibility to unify both dark matter and normal matter in weak isospin multiplets, besides revealing a novel stability mechanism for dark matter with the aid of electric and color charge conservations, since the dark matter is electrically and color neutral \cite{VanLoi:2020kdk,VanLoi:2021dzv}.  

The current experiments have not unraveled any particle content of dark matter. Compared to the rich structure of normal matter within atoms, a structured dark matter is preferred.\footnote{See \cite{Berezhiani:1989fp,Berezhiani:1990sy,Boehm:2003ha,Ma:2006uv,Hur:2007ur,Cao:2007fy} for pioneer works, theoretically/phenomenologically motivated.} Indeed, the simplest possibility---that dark matter contains two types of stable particles---is intriguing, since it may solve the issues of dark matter self-interaction, boosted dark matter, and multiple gamma-ray line \cite{ParticleDataGroup:2020ssz}. In the literature, to stabilize the two components of dark matter simultaneously, a group $Z_2\otimes Z'_2$ has been {\it ad hoc} included as global symmetry. It was indicated that such global symmetry is violated by quantum effects, unless it emerges as a residual gauge symmetry by spontaneous symmetry breaking \cite{Krauss:1988zc}.\footnote{It occurs similarly to continuous global symmetries to be inconsistent with quantum gravity and violated \cite{Mambrini:2015sia}.} In this work, we argue that it is a recognization of dark parities as a result of the dark charge breaking above. This double dark parity leads to three scenarios of cosmological dark matter, two single-component and one two-component schemes. This result has not yet been realized in the literature, specially the last scenario of two-component dark matter will be taken into account under the light of the existing experiments.    

The rest of this work is organized as follows: In Sec. \ref{model}, we give a review of the dark charge and propose double dark parity. In Sec. \ref{scheme} we present three scenarios of dark matter implied by the model. In Sec. \ref{darkmatter}, we examine the case of two-component dark matter, comparing to the observational experiments. Finally, we summarize our results in Sec. \ref{conclusion}.

\section{\label{model} A model of double dark parity}

The standard model is based upon the gauge symmetry $SU(3)_C\otimes SU(2)_L\otimes U(1)_Y $, where the first factor is the QCD group, while the remaining factors are the electroweak symmetry. The electric charge $Q$ always takes the form, 
\be Q = T_3 + Y, \ee
with $T_i$ $(i=1,2,3)$ to be the weak isospin and $Y$ to be the hypercharge. 

It is stressed that the electric charge $Q$  is not quantized, since the $U(1)_Y$ algebra is trivial, i.e. $[Y,Y]=0$, for arbitrary value of $Y$. Further $Y$ is constrained by the anomaly cancelation conditions for the model consistency as well as the gauge-invariant Yukawa Lagrangian for fermion mass generation. But, $Y$ is still arbitrary, since the theory always conserves a hidden symmetry, $N$, which subsequently shifts $Y$ to $Y+x N$, analogous to $B-L$ and $L_i-L_j$ studied in \cite{Babu:1989tq,Babu:1989ex,Foot:1990uf}. 

We add three right-handed neutrinos, $\nu_{aR}$ ($a=1,2,3$), since the left-handed neutrinos might have a nonzero hidden charge, in order for gravity anomaly cancelation mixed with the hidden charge. To obtain the hidden symmetry, we solve the conditions of anomaly cancelation as well as the constraints from Yukawa Lagrangian, for a generic hypercharge $Y$. Assuming $Y(\nu_{aR})=\delta$, such conditions and constraints supply 
\bea && Y(e_{aR})=\delta-1,\hs Y(l_{aL})=\delta-1/2,\crn 
&& Y(u_{aR})=(2-\delta)/3,\hs Y(d_{aR})=-(1+\delta)/3,\hs Y(q_{aL})=(1-2\delta)/6,\eea where $l_{aL}\equiv (\nu_{aL}\ e_{aL})^T$ and $q_{aL}\equiv (u_{aL}\ d_{aL})^T$. The generic hypercharge depends on $\delta$, called the parameter of charge dequantization \cite{VanDong:2020cjf,VanLoi:2020kdk,VanLoi:2021dzv}. 

It is noteworthy that for $\delta=0$, all the particles gain a correct electric charge and hypercharge, so we assign $Q\equiv Q|_{\delta=0}$ and $Y\equiv Y|_{\delta=0}$, as usual. Whereas, for $\delta\ne 0$, all the particles possess a new charge to be a variant of the electric charge and the hypercharge, that defines $D\equiv Q|_{\delta\ne 0}$ and $N\equiv Y|_{\delta\ne 0}$, called dark charge and hyperdark charge, respectively. They are just two solutions according to $\delta=0$ and $\delta\neq 0$. Additionally, the solutions $Y$ and $N$, as well as $Q$ and $D$, are linearly independent, indicating to a novel gauge extension of the standard model, 
\be SU(3)_C\otimes SU(2)_L\otimes U(1)_Y\otimes U(1)_N,\label{gausymmetry} \ee
where $N$ determines $D$, i.e.
\be D = T_3+N, \ee in the same way for the hypercharge and electric charge, $Q=T_3+Y$. Note also that both $Q$ and $D$ neither commute nor close algebraically with the weak isospin. 

The fermion content transforms under the gauge symmetry (\ref{gausymmetry}) as
\bea l_{aL} &=& \left(\begin{array}{c}\nu_{aL}\\ e_{aL}\end{array}\right)\sim \left(1,2,-\fr 1 2, \delta-\fr 1 2\right),\\
\nu_{aR}&\sim & (1,1,0,\delta),\hs e_{aR} \sim  (1,1,-1,\delta-1),\\
q_{aL} &=& \left(\begin{array}{c}u_{aL}\\ d_{aL}\end{array}\right) \sim \left(3,2,\fr 1 6, \fr 1 6-\fr{\delta}{3}\right),\\
u_{aR} &\sim &(3,1, 2/3,2/3-\delta/3), \hs d_{aR} \sim (3,1,-1/3,-\delta/3-1/3),\eea 
where $\delta$ is arbitrarily nonzero, $\delta\neq 0$, and only $\nu_{aR}$ are the new fermions. 

To break the gauge symmetry and produce the masses of the particles properly, the scalar content is given by \be 
\phi=\left(\begin{array}{c}\phi^+\\ \phi^0\end{array}\right)\sim \left(1,2,\fr 1 2, \fr 1 2\right), \hs \chi\sim (1,1,0,-2\delta).\ee Here $\phi$ is the usual Higgs doublet, whose neutral Higgs field, $\phi^0$, is neutral under both the electric charge and dark charge. Hence, the weak vacuum conserves both electric and dark charges. However, the singlet scalar $\chi$ has a nonzero dark charge, $-2\delta$, necessarily presented to break the $U(1)_N$ symmetry, generating appropriate right-handed neutrino masses via the coupling, $\nu_R\nu_R\chi$. It is noteworthy that the dark charge breaking leads to a Majorana neutrino mass $\sim \langle \chi\rangle \nu_R\nu_R$, which constrains the electric charge to be quantized \cite{Babu:1989tq}. Additionally, the left-handed neutrinos couple to the right-handed neutrinos via the Higgs field, hence gain appropriate small masses through the exchange of these heavy right-handed neutrinos \cite{Minkowski:1977sc,GellMann:1980vs,Yanagida:1979as,Glashow:1979nm,Schechter:1980gr}.   

The vacuum expectation values (VEVs) are given by 
\be\langle\phi\rangle= \left(\begin{array}{c}0\\ \fr{v}{\sqrt{2}}\end{array}\right),\hs \langle\chi\rangle=\fr{\Lambda}{\sqrt{2}},\label{vevs}\ee satisfying $\Lambda\gg v=246$ GeV for consistency with the standard model.
The gauge symmetry is broken via two stages, 
\bc \begin{tabular}{c} $SU(3)_C\otimes SU(2)_L\otimes U(1)_Y\otimes U(1)_{N}$ \\
$\downarrow\La$\\
$SU(3)_C\otimes SU(2)_L\otimes U(1)_Y\otimes R_N$\\
$\downarrow v$\\
$SU(3)_C\otimes U(1)_Q\otimes R_D$ \end{tabular}\ec 
where $R_N = e^{ik\pi N/\delta}=(-1)^{kN/\delta}$ for $k$ integer is the intermediate residual symmetry of $U(1)_N$ defined by the new physics scale, $R_N \La=\La$, while \be R_D = e^{ik\pi D/\delta}=(-1)^{kD/\delta}\label{gadttn}
\ee is the final residual symmetry of $U(1)_D$, shifted from $R_N$ by the weak breaking \cite{VanDong:2020cjf,VanLoi:2020kdk,VanLoi:2021dzv}. 

It is clear that if $k=0$, then $R_D=1$, for all fields and every $\delta$, is the identity transformation. To search for the final residual group structure of $R_D$, we find a nonzero minimal value of $|k|$, denoted $m$, that still satisfies $R_D=1$ for all fields. With the $D$ values of the fields in the third column in Table \ref{QDgS} (other stuffs explained below), we derive $m$  dependent on $\delta$ to be
\bea m &=& 2k_1,\\ m/\delta &=& 2k_2,\\ m(\delta-1)/\delta &=& 2k_3,\\ m(\delta-2)/3\delta &=& 2k_4,\\ m(\delta+1)/3\delta &=& 2n,\eea
where $k_{1,2,3,4}$ and $n$ are generically integer (that determine $m$, with a given $\delta$) \cite{VanLoi:2021dzv}. The first equation demands that $m$ must be a positive even integer, i.e. $m=2, 4, 6,\cdots$. The rest yields $k_2 = 3n-m/2, \ k_3 = m-3n, \ k_4 = m/2-2n$ to be integer, as expected, since $m$ is even. Additionally, we obtain the $\delta$ value,
\be\delta = \frac{m}{6n-m}, \ee which also depends the $n$ integer. That said, the residual symmetry $R_D$ is automorphic to an even cyclic group, $\mathcal{Z}_m=\{1,g,g^2,\cdots,g^{m-1}\}$ with $g\equiv (-1)^{D/\delta}$ and $g^m=1$, whose order $m$ is determined via the value of the neutrino dark charge, $\delta$. 

The simplest solution of the residual symmetry corresponds to $m=2$. The corresponding value of $\delta$ that yields such value is 
\be \delta=\fr{1}{3n-1}=-1,1/2,-1/4,\cdots \ee
according to $n=0,\pm 1,\cdots$. The residual symmetry $R_D$ to be automorphic to a dark parity,
\be R_D=\mathcal{Z}_2 = \{1,g\}, \ee
where $g\equiv (-1)^{D(3n-1)}$ and note that $g^2=1$. 

Since the spin parity $h\equiv (-1)^{2s}$ is always conserved by the Lorentz symmetry, we conveniently multiply the residual symmetry with the spin parity group $P_S=\{1,h\}$ to perform $\mathcal{Z}_2\otimes P_S$. This new group has a normal subgroup
\be Z_2 = \{1,p\},  \ee
where \be p \equiv g\times h = (-1)^{D(3n-1)+2s}. \ee
We factorize $\mathcal{Z}_2\otimes P_S=Z_2\otimes [(\mathcal{Z}_2\otimes P_S)/Z_2]$ and note that the quotient group $(\mathcal{Z}_2\otimes P_S)/Z_2=\{Z_2, \{g,h\}\}$ is conserved if $p$, thus $Z_2$, is conserved, because of the $h$ conservation. Hence, we regard $Z_2$ to be the relevant residual symmetry instead of $\mathcal{Z}_2$, i.e. taking $R_D\to Z_2$ into account. 

For comparison, we collect the $p$ values, along with the electric and dark charges, of all fields in Table \ref{QDgS}, where we denote $A$ commonly to be all the gauge fields, except for the $W$ boson. There are two cases for $n$. \ben \item If $n$ is odd, $n\to n_1=\pm 1, \pm 3, \cdots$, then all fields transform trivially under $Z_2$, $p=1$. \item If $n$ is even, $n\to n_2=0, \pm 2, \cdots$, then $\nu$, $u$, $\chi$, $\phi^0$, and $A$ transform as $p=1$, while $e$, $d$, $\phi^+$, and $W^+$ transform as $p=-1$.\een All these cases are presented in Table \ref{QDgS} too, where the (unit) irreducible representation $\underline{1}$ is according to $p=1$, whereas the remaining irreducible representation $\underline{1}'$ is associate to $p=-1$.

\begin{table}[h]
\bc
\begin{tabular}{l|ccccc}
\hline\hline
Field & $Q$ & $D$ & $p$ & $Z_2(n\ \mathrm{odd})$ & $Z_2(n\ \mathrm{even})$\\
\hline
$\nu$ & $0$ & $\delta$ & $1$ & $\underline{1}$ & $\underline{1}$\\
$e$ & $-1$ & $\delta-1$ & $(-1)^{n-1}$ & $\underline{1}$ & $\underline{1}'$\\
$u$ & $2/3$ & $(2-\delta)/3$ & $1$ & $\underline{1}$ & $\underline{1}$\\
$d$ & $-1/3$ & $-(1+\delta)/3$  & $(-1)^{n-1}$ & $\underline{1}$ & $\underline{1}'$\\
$\chi$ & $0$ & $-2\delta$ & $1$ & $\underline{1}$ & $\underline{1}$\\
$\phi^+,W^+$ & $1$ & $1$ & $(-1)^{n-1}$ & $\underline{1}$ & $\underline{1}'$\\
$\phi^0,A$ & $0$ & $0$ & $1$ & $\underline{1}$ & $\underline{1}$\\ 
\hline\hline
\end{tabular}
\caption{$Q$, $D$, $p$ values and $Z_2$ representations for the model fields.}
\label{QDgS}  
\ec
\end{table}

Last, but not least, each value of $\delta$ defines a corresponding $U(1)_N$ factor, since two distinct values of $\delta\to \delta_{1,2}$ lead to two linearly independent solutions for hyperdark charge, $U(1)_N\to U(1)_{N_1}\otimes U(1)_{N_2}$ \cite{VanLoi:2020kdk}. The above two cases may be simultaneously presented in the latter model relevant to the two $U(1)$ factors, if introduced, recognizing a novel double dark parity, $Z_2\otimes Z'_2$, corresponding to $\delta_1=1/(3n_1-1)$ for $n_1$ odd and $\delta_2=1/(3n_2-1)$ for $n_2$ even. It is stressed that within a $U(1)_N$ factor, a residual symmetry $\mathcal{Z}_4$ according to $m=4$ may be hinted; but, this $\mathcal{Z}_4$ is not isomorphic to a $Z_2\otimes Z'_2$ \cite{VanLoi:2021dzv}. The Klein symmetry $Z_2\otimes Z'_2$ widely used, that recognizes two distinct kinds of odd fields of dark matter, might only arise from the two $U(1)$'s symmetry.  

\section{\label{scheme}Schemes of cosmological dark matter}
The first and second scenarios of dark matter are discussed within a single $U(1)_N$ framework, while the third scenario of dark matter necessarily extends $U(1)_N\to U(1)_{N_1}\otimes U(1)_{N_2}$ for viability. 

\subsection{First scenario of single-component dark matter}

According to the first solution above, $\delta\to\delta_1=1/(3n_1-1)$ for $n_1$ to be an odd integer, all the fields in the model transform trivially under the dark parity group $Z_2=Z_2(n\to n_1)$.\footnote{This dark parity transforms similarly to $R$-parity on normal fields, but it differs from that in supersymmetry. As a matter of fact, the dark charge does not commute with the weak isospin, while $B-L$ that defines $R$-parity (cf. e.g. \cite{VanDong:2020bkg,VanDong:2018yae}) does as well as has a nature distinct from the dark charge. Hence, the discrimination of the two kinds of parity is in dark matter implied.}  

Hence, the model provides a natural stability mechanism for single-component dark matter, in which the dark matter candidate transforms nontrivially under $Z_2$, such as
\be \Psi_1\sim \left(1,1,0,\frac{2d_1}{3n_1-1}\right)\sim \underline{1}' \ee 
for a fermion or
\be \Psi_1\sim \left(1,1,0,\frac{2d_1+1}{3n_1-1}\right) \sim \underline{1}'\ee 
for a scalar, where $d_1$ is arbitrarily integer. Hereafter, we also assume all dark matter candidates to be a spin-0 bosonic or spin-1/2 fermionic field and a singlet under the standard model.

Hence, we get the simplest dark matter candidate to be either a fermion or a scalar with $d_1 = 0$ and $n_1=1$. Further, since $Z_2$ is conserved, $\Psi_1$ can obtain an arbitrary mass, which does not decay to the usual particles, responsible for dark matter.

\subsection{Second scenario of single-component dark matter}

For the second solution, $\delta\to\delta_2=1/(3n_2-1)$ for $n_2$ to be an even integer, the model provides a natural stability mechanism for single-component dark matter, where the dark matter candidate, called $\Psi_2$, and all $e$, $d$, $\phi^+$, $W^+$ transform nontrivially under $Z'_2=Z_2(n\to n_2)$, in which 
\be \Psi_2\sim \left(1,1,0,\frac{2d_2}{3n_2-1}\right)\sim \underline{1}' \ee 
for a fermion or
\be \Psi_2\sim \left(1,1,0,\frac{2d_2+1}{3n_2-1}\right)\sim \underline{1}' \ee 
for a scalar, where $d_2$ is arbitrarily integer. Therefore, we obtain the simplest dark matter candidate to be either a fermion or a scalar with $d_2 = 0$ and $n_2=0$. 

It is important to note that $\Psi_2$ can have an arbitrary mass. The symmetries $SU(3)_C$, $U(1)_Q$, and $Z'_2$ jointly suppress the decay of the $\Psi_2$ dark matter, if $\Psi_2$ has a mass larger than the ordinary odd particles $(e, d, \phi^+, W^+)$ \cite{VanLoi:2021dzv, VanLoi:2020kdk}. This is because the dark matter is electrically and color neutral, while the rest of odd fields is not. 

\subsection{Scenario of two-component dark matter}

It is clear that the two solutions of $\delta$ according to $\delta_1$ and $\delta_2$, as well as the solution of $\delta=0$, are linearly independent. Additionally, the model of multi dark charges is viable since the theory is free from all the anomalies, as shown in Appendix of \cite{VanLoi:2020kdk}. Therefore, we obtain a model with the full gauge symmetry, such as
\be SU(3)_C\otimes SU(2)_L\otimes U(1)_Y\otimes U(1)_{N_1}\otimes U(1)_{N_2},\label{fullgausymmetry} \ee
where the hyperdark charges are $N_1=Y|_{\delta=\delta_1}$ and $N_2=Y|_{\delta=\delta_2}$. They determine the relevant dark charges, $D_1=Q|_{\delta=\delta_1}=T_3+N_1$ and $D_2=Q|_{\delta=\delta_2}=T_3+N_2$. Note that the fermion content is remained as before. But, each fermion representation possesses a couple of hyperdark charges $(N_1,N_2)$, while each particle is charged under the dark charges as $(D_1,D_2)$.

The $U(1)_{N_1}\otimes U(1)_{N_2}$ symmetry is broken by the VEVs of two scalar singlets, which transform under  (\ref{fullgausymmetry}), such as \bea \chi_1\sim (1,1,0,-2\delta_1,-2\delta_2)\hs \mathrm{and} \hs \chi_2\sim(1,1,0,0,-2\delta_2),\eea where their VEVs are given by 
\be \langle\chi_1\rangle=\frac{\La_1}{\sqrt2}, \hs \langle\chi_2\rangle=\frac{\La_2}{\sqrt2}, \ee 
satisfying $\La_1, \La_2\gg v=246$ GeV. The field $\chi_1$ necessarily produces the right-handed neutrino masses via the gauge-invariant coupling $\nu_R\nu_R \chi_1$, while $\chi_2$ does not. But, the presence of $\chi_{2}$ is necessary, because this scalar and $\chi_1$ together break the new symmetry properly, and that this breaking leads to the residual symmetry $R_{N_1}\otimes R_{N_2}$, as desirable. The weak breaking shifts this symmetry to $R_{D_1}\otimes R_{D_2}$, where $R_{D_i}=(-1)^{kD_i/\delta_i}$ for $i=1,2$.  

As indicated, choosing $\delta_1=1/(3n_1-1)$ for $n_1$ odd and $\delta_2=1/(3n_2-1)$ for $n_2$ even, the symmetry $R_{D_1}\otimes R_{D_2}$ yields a residual Klein group, $Z_2\otimes Z'_2$, where the parity factors are generated by independent dark charges, \be p_i=(-1)^{D_i(3n_i-1)+2s},\ee for $i=1,2$, respectively, after multiplying the spin parity $h=(-1)^{2s}$ to each group factor. 

Hence, the model provides three kinds of dark fields, $\Xi_{1,2,3}\sim (p_1,p_2)$, implied by $p_{1,2}$ values, as determined in Table \ref{threeDM}, in which $d_{1,2}$ are arbitrarily integer.
\begin{table}[h]
\bc
\begin{tabular}{l|cccc}
\hline\hline
Field & $Z_2\otimes Z'_2$ & Fermion & Scalar\\
\hline
$\Xi_1$ & $(-,+)$ & $\left(1,1,0,\frac{2d_1}{3n_1-1},\frac{2d_2-1}{3n_2-1}\right)$ & $\left(1,1,0,\frac{2d_1+1}{3n_1-1},\frac{2d_2}{3n_2-1}\right)$\\
$\Xi_2$ & $(+,-)$ & $\left(1,1,0,\frac{2d_1-1}{3n_1-1},\frac{2d_2}{3n_2-1}\right)$ & $\left(1,1,0,\frac{2d_1}{3n_1-1},\frac{2d_2-1}{3n_2-1}\right)$\\
$\Xi_3$ & $(-,-)$ & $\left(1,1,0,\frac{2d_1}{3n_1-1},\frac{2d_2}{3n_2-1}\right)$ & $\left(1,1,0,\frac{2d_1+1}{3n_1-1},\frac{2d_2-1}{3n_2-1}\right)$\\
\hline\hline
\end{tabular}
\caption{Three distinct kinds of dark fields implied by $Z_2\otimes Z'_2$.}
\label{threeDM}  
\ec
\end{table}
Since each $Z_2$’s factor provides an independent stability mechanism, the residual group $Z_2\otimes Z'_2$ supplies a natural stability mechanism for structured dark matter of at least two-components. If any two of dark fields $\Xi_{1,2,3}$ are introduced (i.e. omit one of $\Xi_{1,2,3}$), they reveal a scheme of the relevant two-component dark matter. If all the fields $\Xi_{1,2,3}$ are presented, the schemes of dark matter are as follows. For instance, if $\Xi_1$ and $\Xi_2$ are imposed to be the lightest of $\Xi_{1,2,3}$ and that $m_{\Xi_3}>m_{\Xi_1}+m_{\Xi_2}$, the model provides two-component dark matter with $\Xi_{1,2}$. Specially in this case, if $m_{\Xi_3}< m_{\Xi_1}+m_{\Xi_2}$ by contrast, all the fields $\Xi_{1,2,3}$ are realistic dark matter components, i.e. one has a scheme of three-component dark matter. Note that $\Xi_{1,2,3}$ may have a self-interaction with appropriate dark charge choice. Last, but not least, the mentioned dark matter candidates can be stabilized with an arbitrary mass, not necessarily to be smaller than the ordinary odd fields $e$, $d$, $\phi^+$, and $W^+$ \cite{VanLoi:2021dzv, VanLoi:2020kdk}.

\section{\label{darkmatter}Phenomenology of the two-component dark matter}

Among the solutions of dark fields in Table \ref{threeDM}, we obtain the simplest dark fields corresponding to $d_1=d_2=0$ and $n_1=1, n_2=0$, i.e. \be \delta_1=1/2, \hs \delta_2=-1. \ee
Additionally, we consider the model with two-component fermion dark matter by imposing only the first two fields of the third column, by which we relabel \be \Xi_1\to F_1\sim (1,1,0,0,1)\hs \mathrm{and}\hs \Xi_2\to F_2\sim (1,1,0,-1/2,0),\ee for clarity. $F_{1,2}$ have masses to be $m_1$ and $m_2$, respectively. The quantum numbers of $F_{1,2}$ and all other multiplets of the model are supplied in Table \ref{Quantumnumber}. The values of $Q$, $D_1$, and $D_2$ charges and $Z_2\otimes Z'_2$ parities for all fields are collected in Table \ref{QD1D2Z2Z2} for convenience in reading.
\begin{table}[h]
\bc
\begin{tabular}{l|ccccccccccc}
\hline\hline
Multiplet & $l_{aL}$ & $q_{aL}$ & $\nu_{aR}$ & $e_{aR}$ & $u_{aR}$ & $d_{aR}$ & $\phi$ & $\chi_1$ & $\chi_2$ & $F_1$ & $F_2$ \\
\hline
$SU(3)_C$ & $1$ & $3$ & $1$ & $1$ & $3$ & $3$ & $1$ & $1$ & $1$ & $1$ & $1$ \\
$SU(2)_L$ & $2$ & $2$ & $1$ & $1$ & $1$ & $1$ & $2$ & $1$  & $1$  & $1$  & $1$ \\
$Y$ & $-1/2$ & $1/6$ & $0$ & $-1$ & $2/3$ & $-1/3$ & $1/2$ & $0$  & $0$  & $0$  & $0$ \\
$N_1$ & $0$ & $0$ & $1/2$ & $-1/2$ & $1/2$ & $-1/2$ & $1/2$ & $-1$  & $0$  & $0$  & $-1/2$ \\
$N_2$ & $-3/2$ & $1/2$ & $-1$ & $-2$ & $1$ & $0$ & $1/2$ & $2$ & $2$ & $1$ & $0$ \\ \hline\hline
\end{tabular}
\caption{Quantum numbers of the model multiplets.}
\label{Quantumnumber}  
\ec
\end{table}
\begin{table}[h]
\bc
\begin{tabular}{l|cccccccccc}
\hline\hline
Field & $\nu$ & $e$ & $u$ & $d$ & $\chi_1$ & $\chi_2$ & $\phi^+,W^+$ & $\phi^0,A$ & $F_1$ & $F_2$ \\
\hline
$Q$ & $0$ & $-1$ & $2/3$ & $-1/3$ & $0$ & $0$ & $1$ & $0$  & $0$  & $0$  \\
$D_1$ & $1/2$ & $-1/2$ & $1/2$ & $-1/2$ & $-1$ & $0$ & $1$ & $0$  & $0$  & $-1/2$  \\
$D_2$ & $-1$ & $-2$ & $1$ & $0$ & $2$ & $2$ & $1$ & $0$  & $1$  & $0$  \\
$Z_2\otimes Z'_2$ & $(+,+)$ & $(+,-)$ & $(+,+)$ & $(+,-)$ & $(+,+)$ & $(+,+)$ & $(+,-)$ & $(+,+)$  & $(-,+)$  & $(+,-)$  \\ \hline\hline
\end{tabular}
\caption{$Q$, $D_1$, and $D_2$ charges and $Z_2\otimes Z'_2$ parities of the model fields.}
\label{QD1D2Z2Z2}  
\ec
\end{table}

\begin{figure}[h]
\includegraphics[scale=0.9]{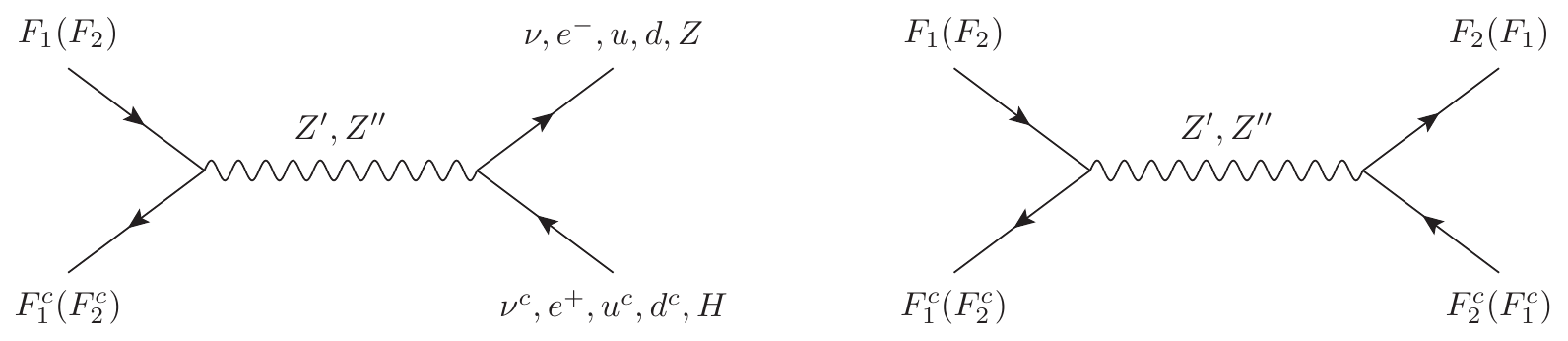}
\caption[]{\label{fig1} Channels for dark matter pair annihilation into standard model particles (left) and conversions between dark matter components (right).}
\end{figure}

We assume that the dark matter components, $F_1$ and $F_2$, have a nature of weakly-interacting massive particles (WIMPs), which are produced through the freezeout mechanism, governed by the $U(1)_{N_1}\otimes U(1)_{N_2}$ gauge portals. [Note that $F_{1,2}$ do not interact with the Higgs fields.] To consider the relic abundance as well as direct detection for the dark matter components, we draw the relevant Feynman diagrams in Fig. \ref{fig1}, which describe dark matter pair annihilation into the standard model particles and the conversion between dark matter components. Here, $Z$ and $H$ are the neutral gauge and Higgs bosons of the standard model, respectively, while $Z'$ and $Z''$ are the new neutral gauge bosons associated with the $U(1)_{N_1}\otimes U(1)_{N_2}$ groups. Let us note that the processes that govern direct dark matter detection signals are given by the $t$-channel diagrams similar to those in the left side of Fig. \ref{fig1}.

First note that the dark matter relic abundance is solved from the coupled Boltzmann equations that describe the yields of $F_{1,2}$. Given that the production of lighter dark matter component from heavier dark matter component is less significant than their annihilation to the standard model particles, we get the approximate solution \cite{Nam:2020twn}
\be \Omega_{F_1}h^2 \simeq \frac{0.1\text{ pb}}{\langle \sigma v\rangle_{F_1}}, \hs \Omega_{F_2}h^2 \simeq \frac{0.1\text{ pb}}{\langle \sigma v\rangle_{F_2}},  \ee
where the thermal average annihilation cross-section times the relative velocity of each  dark matter component is determined by \be \langle \sigma v\rangle_{F_1}=\langle \sigma v\rangle_{F_1F_1\to\text{SM SM}}+\langle \sigma v\rangle_{F_1F_1\to F_2F_2}, \hs \langle \sigma v\rangle_{F_2}=\langle \sigma v\rangle_{F_2F_2\to\text{SM SM}}\ee
if $m_1>m_2$, or \be \langle \sigma v\rangle_{F_1}=\langle \sigma v\rangle_{F_1F_1\to\text{SM SM}}, \hs\langle \sigma v\rangle_{F_2}=\langle \sigma v\rangle_{F_2F_2\to\text{SM SM}}+\langle \sigma v\rangle_{F_2F_2\to F_1F_1}\ee
if $m_2>m_1$. Hence, one has the total dark matter relic abundance to be 
\be \Omega_{\text{DM}}h^2 = \Omega_{F_1}h^2 + \Omega_{F_2}h^2. \ee

Applying the Feynman rules for the diagrams in Fig. \ref{fig1}, we obtain
\bea
\langle \sigma v\rangle_{F_1F_1\to\text{SM SM}} &\simeq & \frac{m_1^2}{16\pi}\sum_{f,i,j}N_C(f)\frac{C_{F_1F_1Z_i}C_{F_1F_1Z_j} [g_V^{Z_i}(f)g_V^{Z_j}(f)+g_A^{Z_i}(f)g_A^{Z_j}(f)]}{(4m_1^2-m_{Z_i}^2)(4m_1^2-m_{Z_j}^2)}\crn
&& + \frac{m_1^2}{16\pi m^2_Z}\sum_{i,j}\frac{C_{F_1F_1Z_i}C_{F_1F_1Z_j} C_{ZHZ_i}C_{ZHZ_j}}{(4m_1^2-m_{Z_i}^2)(4m_1^2-m_{Z_j}^2)},\\
\langle \sigma v\rangle_{F_1F_1\to F_2F_2} &\simeq & \frac{\sqrt{m_1^2-m_2^2}(2m_1^2 + m_2^2)}{2\pi m_1} \sum_{i,j}\frac{C_{F_1F_1Z_i} C_{F_1F_1Z_j}C_{F_2F_2Z_i} C_{F_2F_2Z_j}}{(4m_1^2-m_{Z_i}^2)(4m_1^2-m_{Z_j}^2)},\\
\langle \sigma v\rangle_{F_2F_2\to\text{SM SM}} &=& \langle \sigma v\rangle_{F_1F_1\to\text{SM SM}} (F_1\leftrightarrow F_2, m_1\leftrightarrow m_2),\\
\langle \sigma v\rangle_{F_2F_2\to F_1F_1} &=& \langle \sigma v\rangle_{F_1F_1\to F_2F_2}(F_1\leftrightarrow F_2, m_1\leftrightarrow m_2),
\eea
where $N_C$ is the color number, $Z_i,Z_j = Z', Z''$, and $f$ refers to every standard model fermion. The couplings of $Z'$ with the standard model fermions are supplied in Table \ref{couplings}, while
\be C_{F_1F_1Z'} \simeq -g_2 s_\xi,\hs C_{F_2F_2Z'} \simeq -\fr 1 2 g_1 c_\xi,\hs C_{ZHZ'} \simeq -\fr{1}{2c_W}g v (g_1c_\xi-g_2s_\xi), \ee
in which $g,g_1,g_2$ are the coupling constants according to $SU(2)_L, U(1)_{N_1}, U(1)_{N_2}$ groups, $c_W$ is the cosine of the Weinberg angle. The mixing angle ($\xi$) between the new neutral gauge bosons and their masses are given by 
\bea
t_{2\xi}&\equiv &\tan (2\xi)  \simeq  \frac{4g_1g_2\La_1^2}{g_1^2\La_1^2-4g_2^2(\La_1^2+\La_2^2)},\\
 m^2_{Z',Z''} &\simeq & \frac{1}{2}\left\{g_1^2\La_1^2 + 4g_2^2(\La_1^2+\La_2^2)\mp\sqrt{[g_1^2\La_1^2 - 4g_2^2(\La_1^2+\La_2^2)]^2+16g_1^2g_2^2\La_1^4}\right\}.
\eea
\begin{table}[!h]
\bc
\begin{tabular}{l|cc}
\hline\hline
$f$ & $g^{Z'}_V(f)$ & $g^{Z'}_A(f)$\\ 
\hline
$\nu_a$ & $3 g_2 s_\xi$ & $3 g_2 s_\xi$ \\
$e_a$ & $7g_2s_\xi-g_1c_\xi$ & $g_1c_\xi-g_2s_\xi$ \\
$u_a$ & $g_1c_\xi-3g_2s_\xi$ & $g_2s_\xi-g_1c_\xi$ \\
$d_a$ & $-g_1c_\xi-g_2s_\xi$ & $g_1c_\xi-g_2s_\xi$ \\
\hline\hline
\end{tabular}
\caption{Couplings of $Z'$ with the standard model fermions.}  
\label{couplings}
\ec
\end{table}
Note that the couplings of $Z''$ to the particles can be obtained from those of $Z'$ by replacing $c_\xi\to s_\xi, s_\xi\to -c_\xi$, which need not necessarily be determined.

There is a mixing between $Z$ and $Z',Z''$ which deviates the rho parameter by an amount, 
\be
\Delta\rho\equiv\rho-1=\frac{m^2_W}{c^2_Wm^2_{Z}}\simeq\frac{v^2}{16}\left(\frac{4}{\La^2_1}+\frac{9}{\La^2_2}\right).
\ee
Using the global fit $\Delta\rho \leq 0.00058$ \cite{ParticleDataGroup:2020ssz}, we limit the new physics scales, $\La_1\geq 5.1 \text{ TeV}$ for $\La_1\ll\La_2$, $\La_2\geq 7.7 \text{ TeV}$ for $\La_1\gg\La_2$, and $\La_{1,2}\geq 9.2 \text{ TeV}$ for $\La_1\simeq\La_2$. These limits are appropriate to the $Z$ decay width and various collider bounds studied in \cite{VanLoi:2020kdk} for a single $U(1)_N$, which can be translated to this model without significant change.   

To study the direct detection for dark matter components, we determine the effective Lagrangian describing dark matter-normal matter interactions as induced by the new neutral gauge bosons, 
\bea \mathcal{L}^{\text{eff}}_{F_1} &=& \sum_{i,q}\fr{1}{4m^2_{Z_i}} C_{F_1F_1Z_i}\bar{F}_1\gamma^\mu F_1 \bar{q}\gamma_\mu [g_V^{Z_i}(q)-g_A^{Z_i}(q)\gamma_5]q,\\
\mathcal{L}^{\text{eff}}_{F_2} &=& \mathcal{L}^{\text{eff}}_{F_1} (F_1\leftrightarrow F_2), \eea
where $Z_i=Z',Z''$ and $q=u,d$. Hence, the spin-independent (SI) scattering cross-sections of the dark matter components with target nucleus are given by \cite{Barger:2008qd}
\bea \sigma^{\text{SI}}(F_1) &\simeq& \sum_i \frac{m_N^2}{16\pi m^4_{Z_i}}C_{F_1F_1Z_i}^2\left[g_V^{Z_i}(u)(Z+A)+g_V^{Z_i}(d)(2A-Z)\right]^2,\label{adtt12}\\
\sigma^{\text{SI}}(F_2) &=& \sigma^{\text{SI}}(F_1)(F_1\leftrightarrow F_2),\label{adtt13}
 \eea
where $Z_i=Z',Z''$, $m_N$ is the nucleon mass, and $Z,A$ are the nucleus charge and the total number of nucleons in the nucleus, respectively. Note that the dark matter--nucleon reduced masses do not depend on $m_{1,2}$, since these dark matter masses are much bigger than the nucleon mass (see below). We obtain the effective SI cross-section for each dark matter component as
\bea \sigma^{\text{SI}}_{\text{eff}}(F_1) &=& \frac{\Om_{F_1}h^2}{\Om_{\text{DM}}h^2}\sigma^{\text{SI}}(F_1),\label{adtt14}\\
\sigma^{\text{SI}}_{\text{eff}}(F_2) &=& \frac{\Om_{F_2}h^2}{\Om_{\text{DM}}h^2}\sigma^{\text{SI}}(F_2).\label{adtt15}
\eea
Further, the numerical investigation will use the following parameter values,
\bea && v = 246 \text{ GeV}, \hs s^2_W \simeq 0.231, \\ 
&& g\simeq 0.652, \hs m_Z \simeq 91.187 \text{ GeV}, \\ 
&& Z = 54, \hs A = 131, \hs m_N \simeq 1 \text{ GeV}.\eea

\begin{figure}[h]
\includegraphics[scale=0.37]{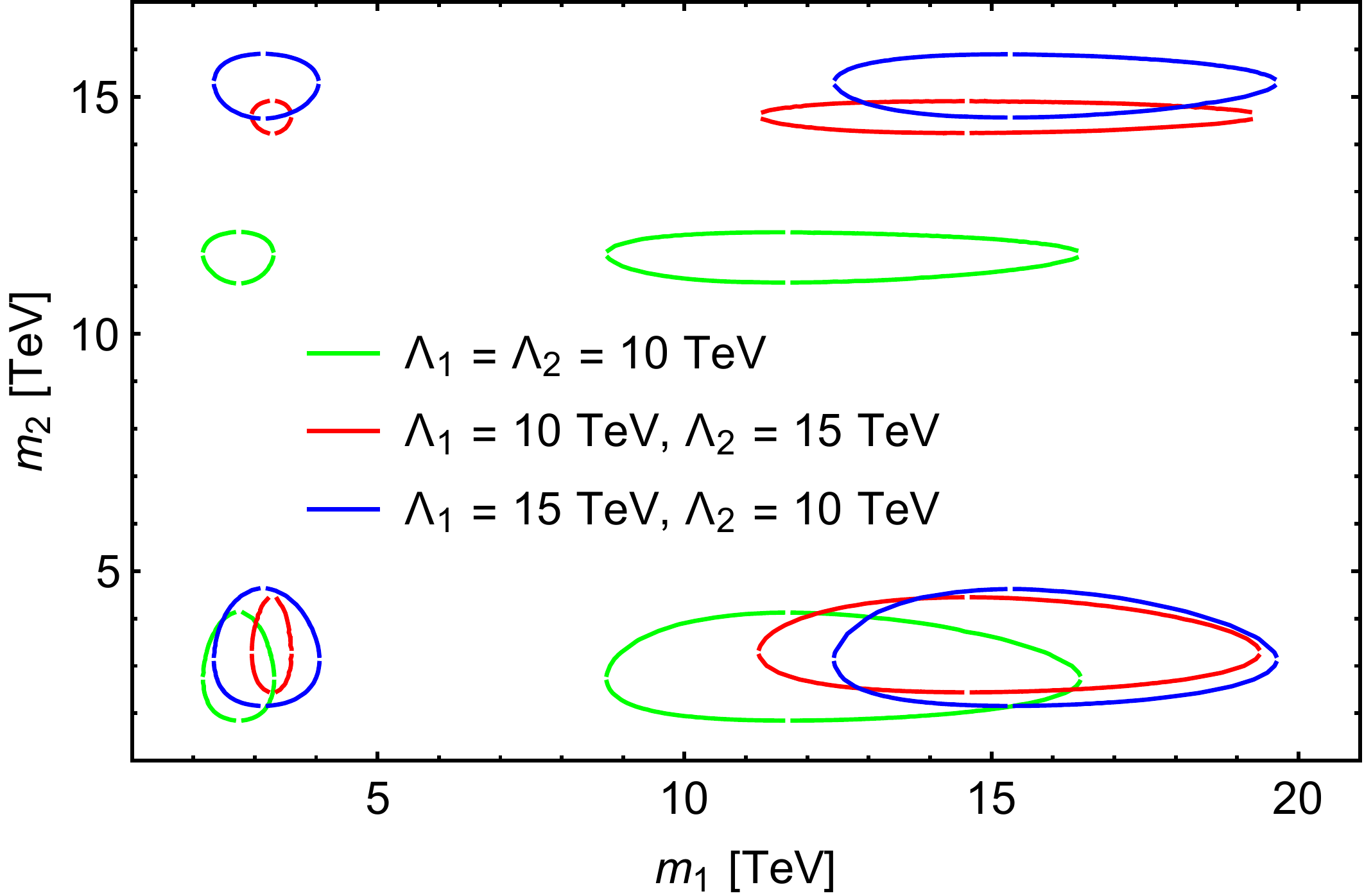}
\includegraphics[scale=0.37]{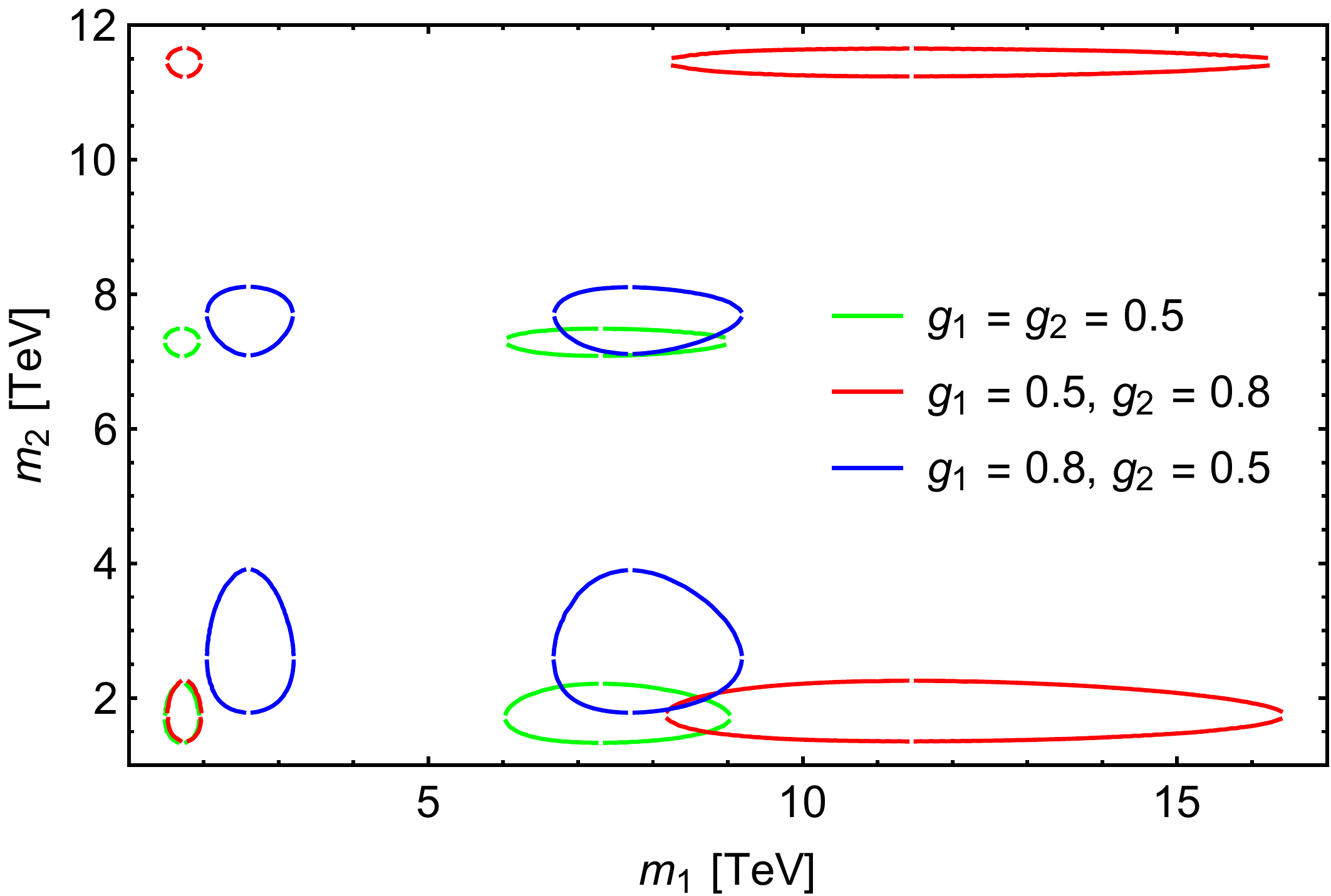}
\caption[]{\label{fig2}Total dark matter relic density contoured as a function of component dark matter masses for different choices of $\La_1,\La_2$ with fixed $g_1=g_2=0.8$ (left panel) and of $g_1,g_2$ with fixed $\La_1=\La_2=10$ TeV (right panel).}
\end{figure}

In Fig. \ref{fig2}, we make contours of the total relic density, $\Om_{\text{DM}}h^2=0.12$ \cite{ParticleDataGroup:2020ssz}, as a function of the dark matter masses, $m_1$ and $m_2$, for $g_1=g_2=0.8$, according to the several choices of $\La_1$ and $\La_2$ (left panel), as well as for $\La_1=\La_2=10$ TeV, according to the several choices of $g_1$ and $g_2$ (right panel). It is clear that the disconnected (very narrow) regions on each curve are due to the $Z',Z''$ mass resonances, $m_1=m_2=m_{Z'}/2$ and $m_1=m_2=m_{Z''}/2$, in the relic density. Note that such a resonance reduces the relic density to zero, so the disconnected regions are omitted for the correct density. Additionally, since the $Z',Z''$ masses are quite separated and that the resonances are strongly derived by $Z',Z''$, the allowed regions of the relic density are not overlapped, resulting as separated, closed curves on the dark matter mass ranges. The figure shows that the dark matter components obtain a mass in the TeV region.

It is clear that the SI cross-sections of $F_{1,2}$ from (\ref{adtt12}) and (\ref{adtt13}) do not depend on the heavy dark matter masses, but on the $Z',Z''$ masses and couplings. In Tab. \ref{results}, we determine these SI cross-sections of $F_{1,2}$ according to the benchmark parameters of $\La_{1,2}$ and $g_{1,2}$ above, and note that they become the measured effective SI cross-sections as in (\ref{adtt14}) and (\ref{adtt15}), given that the dark matter components, $F_1$ and $F_2$, dominate over the total relic density, respectively. Because the values of $\sigma^{\mathrm{SI}}({F_{1,2}})$ from the table are all below the experimental bound \cite{XENON:2017vdw,XENON:2018voc}, the measured $\sigma^{\mathrm{SI}}_{\mathrm{eff}}({F_{1,2}})$ also satisfy such bound, since $\sigma^{\mathrm{SI}}_{\mathrm{eff}}({F_{1,2}})\leq \sigma^{\mathrm{SI}}({F_{1,2}})$.  

\begin{figure}[h]
\includegraphics[scale=0.4]{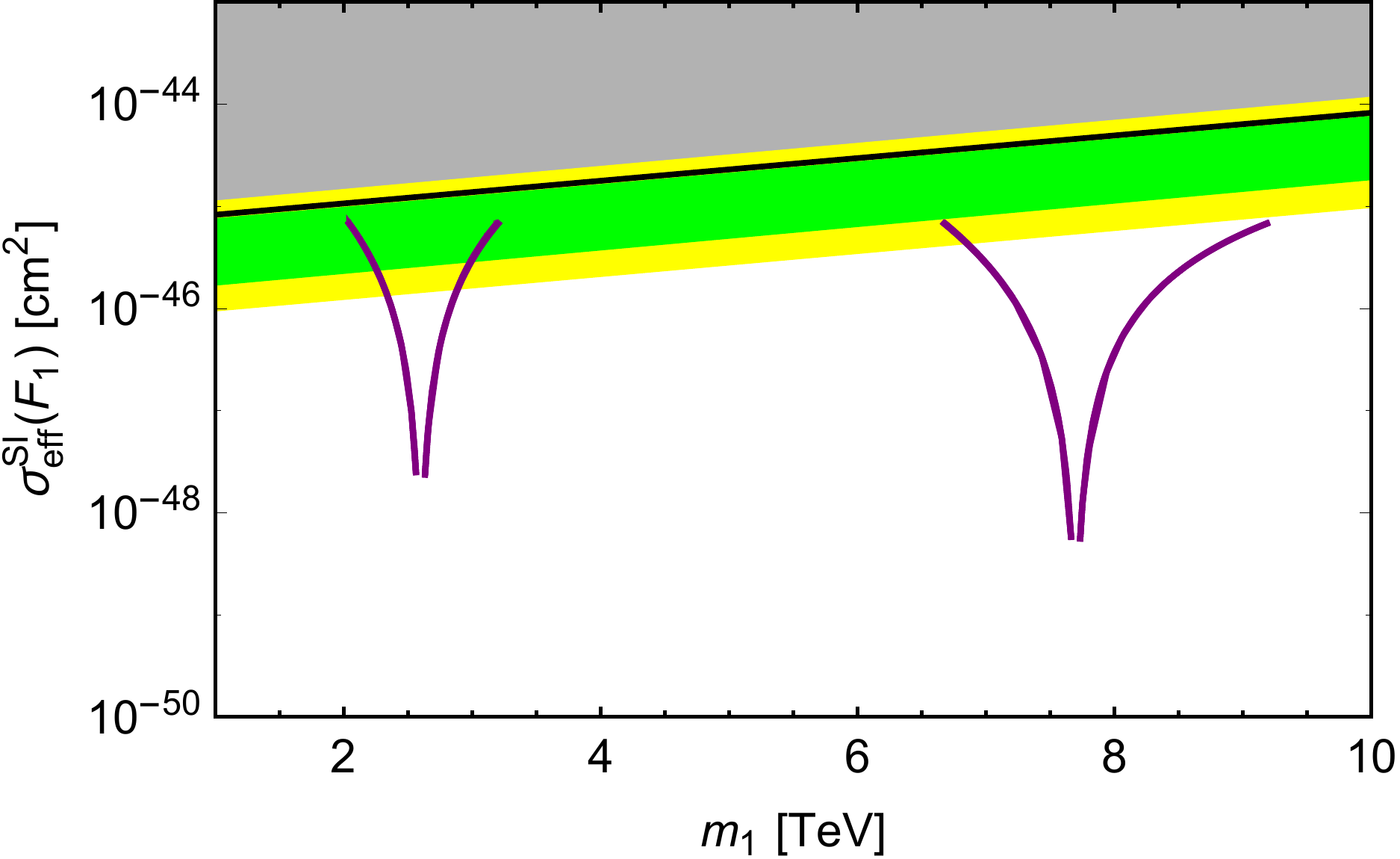}
\includegraphics[scale=0.4]{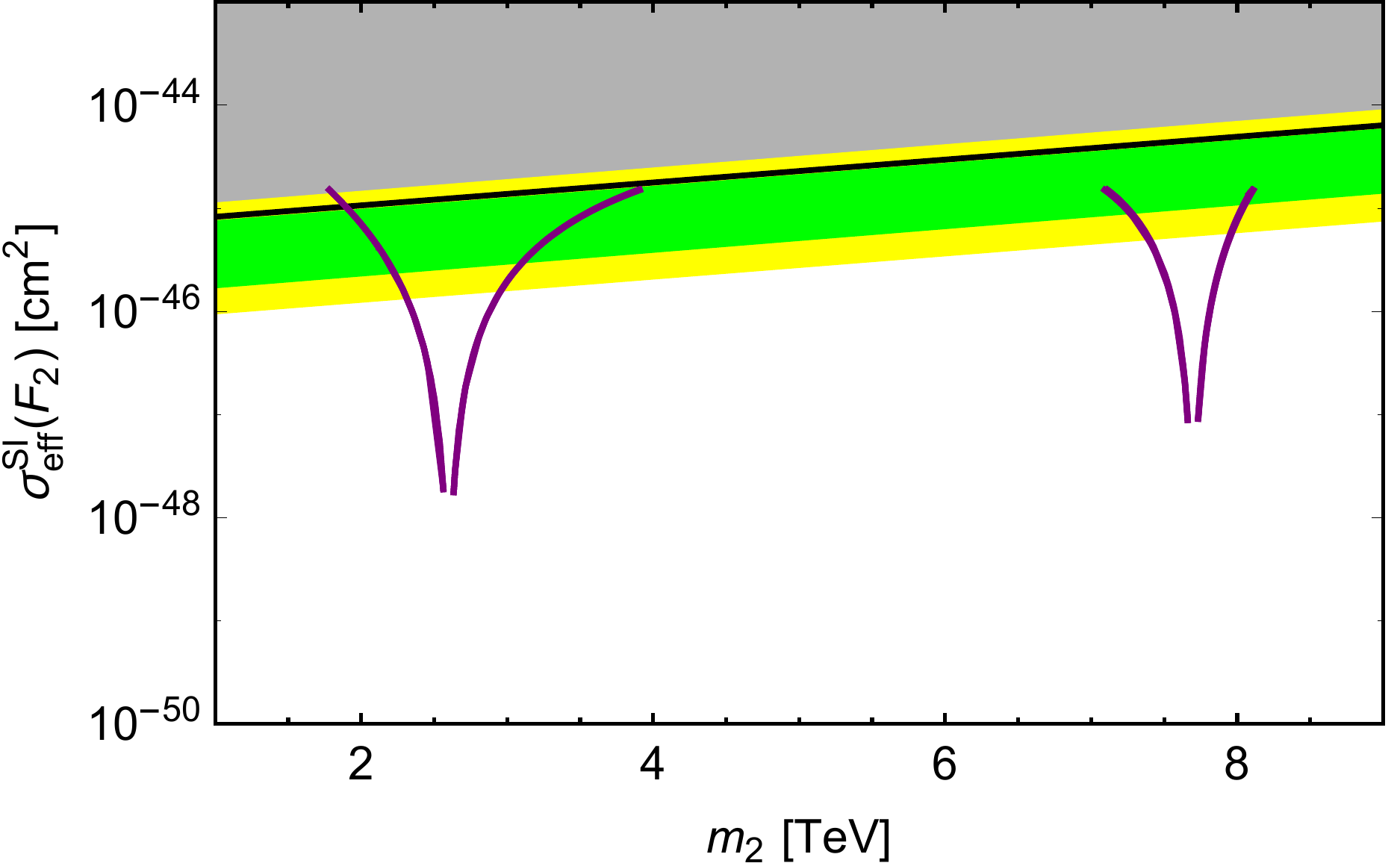}
\caption[]{\label{fig3} Effective SI cross-section of each dark matter component plotted as a function of its mass, for $\La_1=\La_2=10$ TeV and $g_1=0.8$, $g_2=0.5$.}
\end{figure}

Although the effective SI cross-sections agree with the current direct detection, their dependence on the dark matter masses coming only from the contributing factors $\Om_{F_{1,2}}h^2/\Om_{\mathrm{DM}}h^2$ might reduce their values, providing a fit for a future-projected stronger limit of direct detection cross-section. Hence, we will illustrate such a case, for completeness. Supposing that the dark matter components yield a correct total relic density, in Fig \ref{fig3} we plot the effective SI cross-sections for dark matter components as a function of their mass according to $\La_1=\La_2=10$ TeV and $g_1=0.8,\ g_2=0.5$. In this figure, the XENON1T bound has also been shown, with the upper limit (black line), $1\sigma$ (green), and $2\sigma$ (yellow) sensitivity bands. Additionally, the gray space is obviously the excluded region. It is noted that for the benchmark values of $\La_{1,2}$ and $g_{1,2}$, the effective SI cross-sections have a similar shape, but all are below the current bound.  

Since the direct detection is satisfied, we obtain the viable dark matter mass regimes as extracted directly from Fig. \ref{fig2}, which are collected in Table \ref{results}.
\begin{table}[!h]
\bc
\begin{tabular}{cccccccc}
\hline\hline
$g_1$ & $g_2$ & $\La_1$[TeV]& $\La_2$[TeV] &$\sigma^{\text{SI}}_{F_1}/10^{-46}[\text{cm}^2]$ & $\sigma^{\text{SI}}_{F_2}/10^{-45}[\text{cm}^2]$ & $m_1$[TeV] & $m_2$[TeV]\\ \hline
$0.8$ & $0.8$ & $10$ & $10$ & $6.41786$ & $1.27936$ & 2.14--3.32 and 8.71--16.44 & 1.83--4.15 and 11.00--12.17\\
$0.8$ & $0.8$ & $10$ & $15$ & $1.4729$ & $0.21129$ & 2.94--3.61 and 11.19--19.38 & 2.42--4.45 and 14.22--14.92\\
$0.8$ & $0.8$ & $15$ & $10$ & $7.87098$ & $1.32997$ & 2.32--4.05 and 12.42--19.66 & 2.14--4.63 and 14.53--15.91\\
$0.5$ & $0.5$ & $10$ & $10$ & $6.41786$ & $1.27936$ & 1.47--1.95 and 6.05--9.04 & 1.32--2.21 and 7.07--7.49\\
$0.5$ & $0.8$ & $10$ & $10$ & $6.39605$ & $1.18251$ & 1.5--1.98 and 8.17--16.41 & 1.34--2.26 and 11.22--11.66\\
$0.8$ & $0.5$ & $10$ & $10$ & $6.77424$ & $1.50503$ & 2.03--3.21 and 6.67--9.19 & 1.77--3.91 and 7.08--8.12\\
\hline\hline
\end{tabular}
\caption{Benchmark $g_{1,2}$ and $\La_{1,2}$ values, corresponding SI scattering cross-sections, and viable dark matter mass regimes.}  
\label{results}
\ec
\end{table}

\section{\label{conclusion} Conclusion and outlook}

The simplest way to have a multicomponent dark matter scenario adds to the standard model an exact symmetry, $Z_2\otimes Z'_2$. One also adds an exact $Z_2$ symmetry to supersymmetric models, universal extra-dimension models, or $B-L$ models. Supersymmetric models with $\mathcal{N}=2$ also reveal it. A period, the existence of such $Z_2\otimes Z'_2$ symmetry from gauge principle is questioning and doubtful (cf. \cite{Petersen:2009ip} for a discussion). To our best knowledge, a $Z_4$ group is the smallest residual gauge symmetry recognized consistent with multicomponent dark matter \cite{Batell:2010bp,Belanger:2014bga,Yaguna:2019cvp}. In this work, we have shown that each $Z_2$ factor arises from a dark charge symmetry and that the double dark charges are needed to obtain a double dark parity. A model of two-component dark matter recognizing this double dark parity shown obey the observations. Further, the $U(1)$ factors may be well hinted from a GUT, GUT flipped, or string compactification, which are worth exploring.        

\section*{Acknowledgments}

This research is funded by Vietnam National Foundation for Science and Technology Development (NAFOSTED) under grant number 103.01-2019.353.

\bibliographystyle{JHEP}
\bibliography{combine}

\end{document}